\input amstex
\magnification 1200
\TagsOnRight
\def\qed{\ifhmode\unskip\nobreak\fi\ifmmode\ifinner\else
 \hskip5pt\fi\fi\hbox{\hskip5pt\vrule width4pt
 height6pt depth1.5pt\hskip1pt}}
 \def\adots{\mathinner{\mkern2mu\raise1pt\hbox{.}
\mkern3mu\raise4pt\hbox{.}\mkern1mu\raise7pt\hbox{.}}}
\def\sdots{\mathinner{
     \mskip.01mu\raise1pt\vbox{\kern1pt\hbox{.}}
     \mskip.01mu\raise3pt\hbox{.}
     \mskip.01mu\raise5pt\hbox{.}
\mskip1mu}}
\baselineskip 19 pt
\parskip 5 pt

\centerline {\bf TIME EVOLUTION OF THE SCATTERING DATA}
\centerline {\bf FOR A FOURTH-ORDER LINEAR DIFFERENTIAL OPERATOR}

\vskip 10 pt
\centerline {Tuncay Aktosun}
\vskip -6 pt
\centerline {Department of Mathematics}
\vskip -6 pt
\centerline {University of Texas at Arlington}
\vskip -6 pt
\centerline {Arlington, TX 76019-0408, USA}

\centerline{Vassilis G. Papanicolaou}
\vskip -6 pt
\centerline{Department of Mathematics}
\vskip -6 pt
\centerline{National Technical University of Athens}
\vskip -6 pt
\centerline{Zografou Campus, 157 80, Athens, Greece}

\vskip 10 pt

\noindent {\bf Abstract}: The time evolution of the scattering and spectral data
is obtained for the differential operator $\displaystyle\frac{d^4}{dx^4}
+\displaystyle\frac{d}{dx} u(x,t)\displaystyle\frac{d}{dx}+v(x,t),$
where $u(x,t)$ and $v(x,t)$ are real-valued potentials
decaying exponentially as $x\to\pm\infty$ at each fixed $t.$
The result is relevant in a crucial step
of the inverse scattering transform method that is
used in solving the initial-value problem for a pair
of coupled nonlinear partial differential equations
satisfied by $u(x,t)$ and $v(x,t).$

\vskip 15 pt
\par \noindent {\bf Mathematics Subject Classification (2000):}
37K15 35Q51
\vskip -6 pt
\par\noindent {\bf Keywords:} Time evolution of scattering data for a fourth-order ODE,
inverse scattering transform
for a fourth-order ODE, Gelfand-Dickey hierarchy

\vskip -6 pt
\par\noindent {\bf Short title:} Time evolution for a fourth-order ODE
\newpage

\noindent {\bf 1. INTRODUCTION}
\vskip 3 pt

Consider the fourth-order ordinary differential equation
$$\varphi''''+(u\varphi')'+v\varphi=k^4\varphi,\qquad x\in{\bold{R}},\tag 1.1$$
where the prime denotes the derivative with respect to the
independent variable $x,$ and the potentials
$u$ and $v$ are real valued
in such a way that $u,$ $u',$ $v$ are integrable and they decay exponentially or faster
[8,9] as $x\to\pm\infty.$ Let us write (1.1) as
$\Cal L\varphi=k^4\varphi,$
where $\Cal L$ is the linear operator defined as
$$\Cal L:=D^4+DuD+v,\qquad D:=\displaystyle\frac{d}{dx}.\tag 1.2$$
We will consider (1.1) in the sector $\Omega$ in the complex $k$-plane where
$\arg k\in[0,\pi/4]$ with the understanding that we exclude
the point $k=0.$ A complete analysis at $k=0$ for (1.1) does not exist and this deserves
a careful future study. Some partial analysis at $k=0$ is available in [8].
We will omit the analysis
at $k=0$ in this paper.

The differential equation appearing in (1.1) is the canonical equivalent [4,6]
of the Euler-Bernoulli equation, and the former is obtained from the latter
via a transformation of both independent and dependent
variables [4]. Such fourth-order equations arise
in analyzing vibrations of beams, whereas second-order equations
are used in describing vibrations of strings.

We refer the reader to [8,9] and [7,10] for two complementary studies
of (1.1). Under appropriate restrictions on the potentials $u$ and $v,$
Iwasaki [8,9] studied (1.1) by analyzing it in
the sector $\Omega$ under the assumption that
there are no bound states and no spectral or nonspectral singularities.
He obtained [8] various properties of solutions to (1.1)
and formulated [9] the inverse problem of recovery of $u$ and $v$ from some
appropriate scattering data.
Iwasaki posed [9] the inverse scattering problem for (1.1) as a boundary-value
problem where the scattering data consisted of
a reflection coefficient and a connection coefficient specified
on the boundary of $\Omega,$ and he provided the proof of uniqueness
for the solution to that inverse problem.
A special case of (1.1) was studied in [7,10] under the very
restrictive assumption that the reflection and connection coefficients
are all zero.
The corresponding inverse problem was formulated [10] as a Riemann-Hilbert
problem on the whole complex $k$-plane, where the set of scattering data is specified
on the rays $\arg k=(l-1)\pi/4$ for $l=1,2,\dots,8.$
It seems to be the case that the authors of [7,10] have not been aware of [8,9].
Some examples of explicit solutions to (1.1) were
provided in [7,10] under the restriction that the reflection and
connection coefficients are all zero. We should indicate that the terminology
used in [7,10] differs from that used in [8,9]. For the benefit of the reader
the relationships between the
quantities in [7,10] and those in [8,9]
are indicated in Section 5 of our paper.

It is already known [5,10,12] that (1.1) is related to the coupled system of
nonlinear partial differential equations given in (4.2),
which is solvable by the inverse scattering transform method
and is related to the Gelfand-Dickey hierarchy [5].
The time evolution of the scattering and bound-state data forms
a crucial step in that inverse scattering transform.

One of our goals in this paper is to analyze the bound states for (1.1) by analyzing
it in the sector $\Omega$ of
the complex $k$-plane, which was also the domain used in [8,9], and
to obtain the time evolution of the corresponding bound-state data; namely,
the time evolution of the dependency and norming constants.
The time evolution of the bound-state norming constants is given in [10]
by analyzing (1.1) in the entire complex $k$-plane.
Another goal of ours is
to obtain the time evolutions of the
scattering and connection coefficients as well as other coefficients
related to scattering, which is absent in [7,10].
In our paper we also clarify how the relevant
quantities in [7,10] are related to those
of [8,9], and we provide some illustrative examples of explicit
solutions to (1.1) helping to understand the corresponding
scattering and bound states better.

Our paper is organized as follows. In Section 2 we provide the preliminaries
and introduce
the Jost solutions $\psi_+$ and $\psi_-$ as well as the exponential solutions
$\phi_+$ and $\phi_-$ to (1.1), and we present the scattering data
in terms of the spatial asymptotics
of the Jost and exponential solutions. In Section 3 we analyze
the bound states associated with (1.1) and
introduce the dependency constants and norming
constants for each bound state.
In Section 4 we consider the system of integrable nonlinear
partial differential equations satisfied by the time-evolved potentials
$u(x,t)$ and $v(x,t),$ and we obtain the time evolution
of the corresponding scattering and
bound-state data. In Section 5 we explain
how the quantities associated with (1.1) and used in [7,10]
are related to those used in [8,9]. Finally, in Section 6 we provide
some illustrative explicit solutions to (1.1).

\vskip 10 pt
\noindent {\bf 2. PRELIMINARIES}
\vskip 3 pt

Consider the case that the potentials $u$ and $v$ in (1.1) are functions of $x$
alone and do not depend on the parameter $t.$ In Section~4 we will
consider the case where $u$ and $v$ depend on both $x$ and $t.$

As in [8] let us introduce the Jost solutions $\psi_+$ and $\psi_-$
to (1.1) in the sector
$\arg k\in(0,\pi/4)$ with the asymptotics
$$\psi_\pm(k,x)=e^{\pm ikx}[1+o(1)],\qquad x\to\pm\infty,\tag 2.1$$
in such a way that $e^{\mp ikx}\psi_\pm(k,x)$ remains bounded
for all $x\in\bold R.$
Similarly, let us introduce the exponential solutions $\phi_+$ and
$\phi_-$ in the sector
$\arg k\in(0,\pi/4)$ satisfying the asymptotics
$$\phi_\pm(k,x)=e^{\mp kx}[1+o(1)],\qquad x\to\pm\infty.\tag 2.2$$
It follows that these solutions satisfy
the respective integral relations given by
$$\phi_+(k,x)=e^{-kx}+
\displaystyle\frac{1}{4k^4}
\displaystyle\sum_{j=1}^4\int_x^\infty
dy\,[k_j^3u(y)-k_j^2u'(y)+k_jv(y)]e^{k_j(x-y)}\phi_+(k,y)
,$$
$$\aligned
\psi_+(k,x)=e^{ikx}&+
\displaystyle\frac{1}{4k^4}
\displaystyle\sum_{j=1,2,4}\int_x^\infty
dy\,[k_j^3u(y)-k_j^2u'(y)+k_jv(y)]e^{k_j(x-y)}\psi_+(k,y)
\\
&-
\displaystyle\frac{1}{4k^4}
\int_{-\infty}^x
dy\,[k_3^3u(y)-k_3^2u'(y)+k_3v(y)]e^{k_3(x-y)}\psi_+(k,y)
,\endaligned$$
$$\phi_-(k,x)=e^{kx}-
\displaystyle\frac{1}{4k^4}
\displaystyle\sum_{j=1}^4\int_{-\infty}^x
dy\,[k_j^3u(y)-k_j^2u'(y)+k_jv(y)]e^{k_j(x-y)}\phi_-(k,y)
,$$
$$\aligned
\psi_-(k,x)=e^{-ikx}&+\displaystyle\frac{1}{4k^4}
\int_x^\infty
dy\,[k_1^3u(y)-k_1^2u'(y)+k_1v(y)]e^{k_1(x-y)}\psi_-(k,y)
\\
&-\displaystyle\frac{1}{4k^4}\displaystyle\sum_{j=2,3,4}\int_{-\infty}^x
dy\,[k_j^3u(y)-k_j^2u'(y)+k_jv(y)]e^{k_j(x-y)}\psi_-(k,y),
\endaligned$$
where we have defined
$$k_1:=k, \qquad k_2:=ik, \qquad k_3:=-k, \qquad k_4:=-ik.\tag 2.3$$

The spatial asymptotics of the Jost solutions and the exponential solutions
can be obtained with the help of
the four integral relations given above. Below, we list those
asymptotics, through which we introduce the scattering and connection
coefficients for (1.1).
For $\arg k\in(0,\pi/4)$ we have
$$\cases
\psi_\pm(k,x)=e^{\pm ikx}[1+o(1)],\qquad x\to\pm\infty,\\
\noalign{\medskip}
\psi_\pm(k,x)=e^{\pm ikx}\left[\displaystyle\frac{1}{T(k)}+o(1)
\right],\qquad x\to\mp\infty,\\
\noalign{\medskip}
\phi_\pm(k,x)=e^{\mp kx}[1+o(1)],\qquad x\to\pm\infty,\\
\noalign{\medskip}
\phi_\pm(k,x)=e^{\mp kx}\left[A(k)+o(1)
\right],\qquad x\to\mp\infty.\endcases\tag 2.4$$
For $\arg k=0$ we have
$$\cases
\psi_\pm(k,x)=e^{\pm ikx}[1+o(1)],\qquad x\to\pm\infty,\\
\noalign{\medskip}
\psi_\pm(k,x)=e^{\pm ikx}\left[\displaystyle\frac{1}{T(k)}+
\displaystyle\frac{R_\pm(k)}{T(k)}e^{\mp 2ikx}+
o(1)
\right],\qquad x\to\mp\infty,\\
\noalign{\medskip}
\phi_\pm(k,x)=e^{\mp kx}[1+o(1)],\qquad x\to\pm\infty,\\
\noalign{\medskip}
\phi_\pm(k,x)=e^{\mp kx}\left[A(k)+o(1)
\right],\qquad x\to\mp\infty.\endcases\tag 2.5$$
For $\arg k=\pi/4$ we have
$$\cases
\psi_\pm(k,x)=e^{\pm ikx}[1+C_\pm(k)e^{\mp kx\mp ikx}
+o(1)],\qquad x\to\pm\infty,\\
\noalign{\medskip}
\psi_\pm(k,x)=e^{\pm ikx}\left[\displaystyle\frac{1}{T(k)}+
o(1)
\right],\qquad x\to\mp\infty,\\
\phi_\pm(k,x)=e^{\mp kx}[1+o(1)],\qquad x\to\pm\infty,\\
\noalign{\medskip}
\phi_\pm(k,x)=e^{\mp kx}\left[A(k)+
B_\pm(k)e^{\pm kx\pm ikx}+o(1)
\right],\qquad x\to\mp\infty.\endcases\tag 2.6$$

We emphasize that
$A(k)$ and $T(k)$ are defined for
$\arg k\in[0,\pi/4],$ $R_+(k)$ and $R_-(k)$ are defined for
$\arg k=0,$ and the four coefficients
$B_+(k),$ $B_+(k),$ $C_+(k),$ and $C_-(k)$ are defined only for
$\arg k=\pi/4.$ The coefficient $T$ is known as
the transmission coefficient, $R_+$ and $R_-$ are
the left and right reflection coefficients, respectively,
$C_+$ and $C_-$ are known as the connection coefficients,
and $B_+$ and $B_-$ are some
coefficients that can be expressed in terms of $A,$ $C_+,$ and $C_-,$
as we will see.
It is either known [8] or can easily be shown that
for $\arg k=0$ we have
$$1+\displaystyle\frac{|R_\pm(k)|^2}{|T(k)|^2}=\displaystyle\frac{1}{|T(k)|^2}
,\qquad \displaystyle\frac{R_\pm(k)}{T(k)}=-\displaystyle\frac{R_\mp(k)^*}{T(k)^*},\qquad
A(k)=A(k)^*,\tag 2.7$$
and for $\arg k=\pi/4$ we have
$$\cases
B_\mp(k)=-iB_\pm(k)^*,\qquad
C_-(k)=\displaystyle\frac{iT(k)^*C_+(k)^*}{T(k)},\\
\noalign{\medskip}
|B_-(k)|=|B_+(k)|,\qquad
|C_-(k)|=|C_+(k)|,\\
\noalign{\medskip}
\displaystyle\frac{1}{T(k)}=A(k)^*+B_\pm(k)C_\pm(k),
\qquad
B_\pm(k)^*+A(k)C_\pm(k)=0,\\
\noalign{\medskip}
\displaystyle\frac{A(k)}{T(k)}=|A(k)|^2-|B_\pm(k)|^2
=|A(k)|^2\left(1-|C_\pm(k)|^2\right),
\endcases\tag 2.8$$
where the asterisk denotes complex conjugation.

\vskip 10 pt
\noindent {\bf 3. BOUND STATES}
\vskip 3 pt

Since the coefficient of the third derivative in (1.1) is zero, it follows from the general
theory of ordinary differential equations that the Wronskian of
any four solutions to (1.1) is independent of $x,$ and that Wronskian is zero
if and only if those four solutions are linearly dependent.
Recall that a Wronskian is defined with the help
of a determinant. For example, the Wronskian involving the Jost and
exponential solutions is given by
$$W_4[\psi_+,\psi_-,\phi_+,\phi_-]:=
\left|\matrix \psi_+&\psi_-&\phi_+&\phi_-\\
\noalign{\medskip}
\psi_+'&\psi_-'&\phi_+'&\phi_-'\\
\noalign{\medskip}
\psi_+''&\psi_-''&\phi_+''&\phi_-''\\
\noalign{\medskip}
\psi_+'''&\psi_-'''&\phi_+'''&\phi_-'''\endmatrix\right|.$$
Using (2.4)-(2.6) we obtain
$$W_4[\psi_+(k,x),\psi_-(k,x),\phi_+(k,x),\phi_-(k,x)]=-16ik^6\displaystyle\frac{A(k)}{T(k)}
,\qquad \arg k\in[0,\pi/4].\tag 3.1$$

The linear independence and boundedness properties of various solutions
to (1.1) help to identify bound-state solutions.
Recall that eigenfunctions of $\Cal L$ correspond to square-integrable
solutions to (1.1), which are also known as bound-state solutions.
It is easy to verify that $\Cal L$
is selfadjoint
and hence its eigenvalues can occur only for real values of $k^4,$
i.e. when $k$ lies on the boundary of the region $\Omega$ introduced in Section 1.
Thus, any positive
eigenvalue of $\Cal L$ can occur only on the ray $\arg k=0$ and any
negative eigenvalue can occur only on the ray
$\arg k=\pi/4.$ If $A(k)=0$ at some $k$-value on the
boundary of the region
$\Omega$ and a square-integrable solution to (1.1) at that $k$-value does not exist,
then we call that $k$-value a spectral singularity of (1.1).
If $A(k)=0$ at some $k$-value in the interior of
$\Omega,$ then we call that $k$-value a nonspectral singularity of (1.1).
By a singularity we refer
to either a spectral or nonspectral singularity. It is already known that
at a singular point the two integral relations given in Section~2 for
$\psi_+(k,x)$ and $\psi_-(k,x),$ respectively, are not solvable [8].
Spectral and nonspectral
singularities for (1.1) may exist, and some explicit examples are
illustrated in Section 6.

Note that a bound state in the region $\Omega$ can occur
only when $A(k)/T(k)=0$ somewhere on the ray $\arg k=0$ or
$\arg k=\pi/4.$ Otherwise, as seen from (3.1), the four solutions
$\psi_+,$ $\psi_-,$ $\phi_+,$ and $\phi_-$ are linearly independent, and
the asymptotics of those four solutions given in (2.5) and (2.6) indicate that
no linear combination of them
can decay simultaneously both as $x\to+\infty$ and $x\to-\infty.$

If there is
a bound state at $k=\kappa$
on the ray $\arg k=0,$ then we must have $A(\kappa)=0$.
This follows from the first identity in (2.7)
implying that $1/T(\kappa)\ne 0$ and the fact that
$A(\kappa)/T(\kappa)=0$ at the bound state.

Since four linearly independent solutions to (1.1) must have respective
asymptotics proportional to $e^{\kappa x}[1+o(1)],$
$e^{-\kappa x}[1+o(1)],$ $e^{i\kappa x}[1+o(1)],$
$e^{-i\kappa x}[1+o(1)]$ as $x\to+\infty,$ and appropriately similar asymptotics
as $x\to-\infty,$ it follows that a bound state at $k=\kappa$
must decay exponentially as both as $x\to+\infty$ and $x\to-\infty.$
In this case, we see from (2.5)
that $\psi_+(\kappa,x)$ and $\psi_-(\kappa,x)$
are two linearly independent solutions to (1.1) and they do not vanish
simultaneously both as $x\to+\infty$ and $x\to-\infty.$
Thus, from (2.5) we conclude that
a bound-state eigenfunction
$\varphi(\kappa,x)$ must be in the form
$$\varphi(\kappa,x)=d_1\phi_+(\kappa,x)=
d_2\phi_-(\kappa,x),\tag 3.2$$
for some nonzero constants $d_1$ and $d_2.$
Since any constant multiple of an eigenfunction is still an
eigenfunction of $\Cal L,$ only the ratio $d_2/d_1$ is relevant and we can call
it a dependency constant at $k=\kappa,$ i.e.
$$\eta(\kappa):=\displaystyle\frac{\phi_+(\kappa,x)}{\phi_-(\kappa,x)},\tag 3.3$$
where $\eta(\kappa)$ is the dependency constant at $k=\kappa$ on the ray
$\arg k=0.$
Defining the bound-state norming constants $d_+$ and $d_-$ as
$$d_\pm(\kappa):=\left[\int_{-\infty}^\infty dx\,\phi_\pm(\kappa,x)
\phi_\pm(\kappa,x)^*\right]^{-1/2},\tag 3.4$$
we see that $d_-=\eta d_+$ and that
$d_\pm(\kappa)\,\phi_\pm(\kappa,x)$ is a normalized
bound-state eigenfunction of the operator $\Cal L.$

Having clarified the status of bound states on the ray
$\arg k=0,$ let us now consider the bound states on
the ray $\arg k=\pi/4.$ If there is
a bound state at $k=\kappa$
on the ray $\arg k=\pi/4,$ then we have three possibilities:

\item{(i)} $A(\kappa)=0$ and $1/T(\kappa)\ne 0.$
In this case, an argument similar to the case
given on the ray $\arg k=0$ shows that
$\phi_+(\kappa,x)$ and $\phi_-(\kappa,x)$ are linearly dependent and
the bound state is simple and has the form given in (3.2).
Recall that a bound state
occurring at $k=\kappa$ is simple if there is only
one linearly independent square-integrable solution to
(1.1) when $k=\kappa.$

\item{(ii)} $A(\kappa)\ne 0$ and $1/T(\kappa)=0.$
In this case, a similar argument indicates that
a bound-state eigenfunction must have the form
$$\varphi(\kappa,x)=c_1\psi_+(\kappa,x)=
c_2\psi_-(\kappa,x),$$
for some nonzero constants $c_1$ and $c_2.$
Since a bound-state eigenfunction is defined up to a constant multiple,
only the ratio $c_2/c_1$ is relevant and we can call
that ratio a dependency constant at $k=\kappa,$ i.e.
$$\gamma(\kappa):=\displaystyle\frac{\psi_+(\kappa,x)}{\psi_-(\kappa,x)},\tag 3.5$$
where $\gamma(\kappa)$ is the dependency constant at $k=\kappa$ on the ray
$\arg k=\pi/4.$
Defining the bound-state norming constants $c_+$ and $c_-$ as
$$c_\pm(\kappa):=\left[\int_{-\infty}^\infty dx\,\psi_\pm(\kappa,x)
\psi_\pm(\kappa,x)^*\right]^{-1/2},\tag 3.6$$
we see that $c_-=\gamma c_+$ and that
$c_\pm(\kappa)\,\psi_\pm(\kappa,x)$ is a normalized
bound-state eigenfunction of the operator $\Cal L.$
In this case there is only one linearly independent bound-state
eigenfunction at $k=\kappa,$ and hence the bound state is simple.

\item{(iii)} $A(\kappa)=0$ and $1/T(\kappa)=0.$ In this case,
$\psi_+(\kappa,x)$ and $\psi_-(\kappa,x)$ are linearly dependent
solutions decaying exponentially as $x\to +\infty$
as well as $x\to -\infty,$ and
$\phi_+(\kappa,x)$ and $\phi_-(\kappa,x)$ are
also two linearly
dependent
solutions decaying exponentially as $x\to +\infty$
as well as $x\to -\infty.$
On the other hand, $\psi_+(\kappa,x)$ and $\phi_+(\kappa,x)$ are linearly independent,
and hence they form a basis for the two-dimensional
eigenspace consisting of bound-state solutions
at $k=\kappa.$ In other words, the multiplicity of
the bound state in this case is two.
The corresponding dependency constants
$\eta(\kappa)$ and $\gamma(\kappa)$ and
the norming constants $c_+(\kappa),$ $c_-(\kappa),$
$d_+(\kappa),$ $d_-(\kappa)$
are defined as in (3.3)-(3.6).

\vskip 10 pt
\noindent {\bf 4. TIME EVOLUTION OF SCATTERING AND OTHER COEFFICIENTS}
\vskip 3 pt

In the previous sections we have considered the Jost and exponential
solutions, the scattering and connection coefficients, and
the bound states associated with (1.1)
in the case where the potentials $u$ and $v$ are functions of $x$ alone.
Let us now assume that the potentials $u$ and $v$ appearing in
(1.1) also depend on the extra parameter $t.$ In that case
all the relevant quantities associated with (1.1) may
also depend on $t$ as well.
In this section we analyze such a dependence on $t$ by interpreting
$t$ as the time variable.

Now consider the time evolution of the potentials
$u(x,t)$ and $v(x,t)$ from their initial values $u(x,0)$ and $v(x,0),$ respectively,
so that the time-evolved linear
operator
$\Cal L$ corresponding to (1.2) is given by
$$\Cal L:=\partial_x^4+\partial_x u(x,t)\partial_x+v(x,t),\tag 4.1$$
where $\partial_x=\partial/\partial x,$ and let [5,10,12]
$$\Cal A:=-8\partial_x^3-6u(x,t)\partial_x-3u_x(x,t),$$
so that $\Cal L$ and $\Cal A$ form a Lax pair.
As easily verified, the differential operator
$\Cal L_t+\Cal L\Cal A-\Cal A\Cal L$ reduces to a scalar multiplication
operator, and in fact we get
$$\Cal L_t+\Cal L\Cal A-\Cal A\Cal L=0,$$
which is equivalent to the system of
nonlinear evolution equations
$$\cases u_t=10u_{xxx}+6uu_x-24v_x,\\
\noalign{\medskip}
v_t=3u_{xxxxx}+3uu_{xxx}+3u_xu_{xx}-6uv_x-8v_{xxx}.\endcases\tag 4.2$$
Note that we use subscripts to denote the appropriate partial derivatives.

Since $u$ and $v$ vanish as $x\to\pm\infty$ at each fixed $t,$
we have  $\Cal A\to -8\partial_x^3$ as $x\to\pm\infty.$ From
(4.1) we also see that
$$\Cal L_t=u_t\partial_x^2+u_{xt}\partial_x+v_t.$$
In order to solve the initial-value problem related to (4.2), i.e. to determine
$u(x,t)$ and $v(x,t)$ that solve (4.2) when $u(x,0)$ and $v(x,0)$ are
specified, we are interested in analyzing the
time evolutions of the scattering and other coefficients associated
with (1.1).

Towards our goal, we first analyze the time evolutions of the Jost solutions
$\psi_+(k,x,t)$ and $\psi_-(k,x,t)$ and the exponential solutions
$\phi_+(k,x,t)$ and $\phi_-(k,x,t),$ from which
the time evolutions of other relevant coefficients are easily extracted.

\noindent {\bf Theorem 4.1} {\it In the region $\arg k\in[0,\pi/4]$ the time
evolutions of $\psi_+(k,x,t),$ $\psi_-(k,x,t),$
$\phi_+(k,x,t),$ and $\phi_-(k,x,t)$ are given by}
$$[\partial_t-\Cal A]\psi_\pm=\mp8ik^3\psi_\pm,\qquad
[\partial_t-\Cal A]\phi_\pm=\mp8k^3\phi_\pm.\tag 4.3$$
{\it Moreover,
the time evolutions of various coefficients appearing in (2.4)-(2.6) are
given by}
$$T(k,t)=T(k,0),\quad A(k,t)=A(k,0),\qquad \arg k\in[0,\pi/4],$$
$$R_\pm(k,t)=R_\pm(k,0)e^{\mp 16ik^3 t},\qquad \arg k=0,$$
$$B_\pm(k,t)=B_\pm(k,0)e^{\pm(8ik^3-8k^3)t},
\qquad C_\pm(k,t)=C_\pm(k,0)e^{\mp(8ik^3-8k^3)t},
\qquad \arg k=\pi/4.$$

\noindent PROOF: The proofs in (4.3) can all be given as in the
case of the time evolution of
$\psi_+$ for $\arg k=\pi/4,$ which is outlined below.
It is known [1-3,11] that
$[\partial_t-\Cal A]\psi_+$ must satisfy $\Cal L
\varphi=k^4\varphi,$ where $\Cal L$ is the operator in (4.1).
Thus, we have
$$[\partial_t-\Cal A]\psi_+=
c_1(k,t)\psi_++c_2(k,t)\psi_-+c_3(k,t)\phi_++c_4(k,t)\phi_-,\tag 4.4$$
for some coefficients $c_j(k,t)$ to be determined.
By evaluating (4.4) as $x\to-\infty$ and $x\to+\infty,$ with
the help of (2.6) we obtain
$$\left(\displaystyle\frac{e^{ikx}}{T}\right)_t-8ik^3
\displaystyle\frac{e^{ikx}}{T}=c_1\displaystyle\frac{e^{ikx}}{T}+c_2[e^{-ikx}+C_-e^{kx}]+
c_3[Ae^{-kx}+B_+e^{ikx}]+c_4e^{kx},\tag 4.5$$
$$\aligned
(C_+)_te^{-kx}-&8ik^3e^{ikx}-8k^3C_+e^{-kx}\\
&=c_1[e^{ikx}+C_+e^{-kx}]+c_2
\displaystyle\frac{e^{-ikx}}{T}+c_3e^{-kx}+c_4[Ae^{kx}+B_-e^{-ikx}].
\endaligned\tag 4.6$$
By matching the corresponding coefficients
of the exponential terms in (4.5) and (4.6) we get
$$c_1=-8ik^3,\qquad c_2=c_3=c_4=0,
\qquad T_t=0, \qquad (C_+)_t=(8k^3-8ik^3)C_+,$$
and hence the first equation in (4.3) for $\psi_+$ is confirmed
when $\arg k=\pi/4,$ and
we also get the time evolutions of $T$ and $C_+$
when $\arg k=\pi/4,$ as stated. The remaining parts of the
proof are obtained in a similar way. \qed

The implication of Theorem 4.1 that $T(k,t)$ and $A(k,t)$ do not
change in $t$ is significant. As we have seen in Section 3,
at a bound state $k=\kappa$ we must have $A(\kappa,t)/T(\kappa,t)=0,$
and at a singularity $k=\kappa$ we must have $A(\kappa,t)=0.$
Hence, the $k$-values corresponding
to bound states or singularities of the operator
$\Cal L$ of (4.1) also remain unchanged
in time.

\noindent {\bf Theorem 4.2} {\it Assume that
$k=\kappa$ corresponds to a bound state of
(1.1). The time evolution of
the bound-state dependency constants $\gamma(\kappa,t)$ and $\eta(\kappa,t)$
and the evolution of the norming constants
$c_\pm(\kappa,t)$ and $d_\pm(\kappa,t)$ are given by}
$$c_\pm(\kappa,t)=c_\pm(\kappa,0)\,e^{\pm 4(1+i)\kappa^3t},\qquad
d_\pm(\kappa,t)=d_\pm(\kappa,0)\,e^{\pm 4(1+i)\kappa^3t},\tag 4.7$$
$$\gamma(\kappa,t)=\gamma(\kappa,0)\,e^{-8(1+i)\kappa^3t},\qquad
\eta(\kappa,t)=\eta(\kappa,0)\,e^{-8(1+i)\kappa^3t}.\tag 4.8$$

\noindent PROOF: Let us assume that there is a bound state at
$k=\kappa$ somewhere on $\arg k=\pi/4$
with $1/T(\kappa,0)=0.$ Then, $\psi_+(\kappa,x,t)$ is a
bound-state solution and the norming constant $c_+(\kappa,t)$ can be defined
as in (3.6) via
$$c_+(\kappa,t):=\left[\int_{-\infty}^\infty dx\,\psi_+(\kappa,x,t)
\psi_+(\kappa,x,t)^*\right]^{-1/2},\tag 4.9$$
so that $c_+(\kappa,t)\psi_+(\kappa,x,t)$ is normalized, i.e. its $L^2$-norm
is equal to one. Let us now find the time evolution of $c_+(\kappa,t).$ From (4.3)
and its complex conjugate we obtain
$$[\partial _t+8\partial_x^3+6\partial_x u(x,t)+3u_x(x,t)]\psi_+(\kappa,x,t)
=-8i\kappa^3\psi_+(\kappa,x,t),\tag 4.10$$
$$[\partial _t+8\partial_x^3+3\partial_x u(x,t)+3u_x(x,t)
]\psi_+(\kappa,x,t)^*
=8i(\kappa^*)^3\psi_+(\kappa,x,t)^*,\tag 4.11$$
where we recall that the potentials $u$ and $v$ are assumed to be real valued.
Multiplying (4.10) by $\psi_+(\kappa,x,t)^*$ and
(4.11) by $\psi_+(\kappa,x,t),$ and adding the resulting equations
we obtain
$$\partial_t |\psi_+|^2+\partial_x\left[8\psi_+^*(\partial_x^2
\psi_+)+8\psi_+ (\partial_x^2\psi_+^*)-
8(\partial_x\psi_+)(\partial_x\psi_+^*)
+6u|\psi_+|^2
\right]
=-8(1+i)\kappa^3|\psi_+|^2,$$
where we have used the fact that $k^*=-ik$ on the ray $\arg k=\pi/4.$
Integrating over the real axis and using the vanishing of
$\psi_+,$ $\partial_x\psi_+,$
$\partial_x^2\psi_+,$ and $u$ as $x\to+\infty$ and $x\to-\infty,$ we
obtain
$$\displaystyle\frac{d}{dt}\int_{-\infty}^\infty
dx\,|\psi_+(\kappa,x,t)|^2=
-8(1+i)\kappa^3\int_{-\infty}^\infty
dx\,|\psi_+(\kappa,x,t)|^2.\tag 4.12$$
Using (4.9) we can write (4.12) as
$$\displaystyle\frac{d}{dt}\left[\displaystyle\frac{1}{c_+(\kappa,t)^2}\right]=
\displaystyle\frac{-8(1+i)\kappa^3
}{c_+(\kappa,t)^2},$$
or equivalently we obtain
$$\displaystyle\frac{d c_+(\kappa,t)}{dt}=
4(1+i)\kappa^3c_+(\kappa,t),$$
which yields
$$c_+(\kappa,t)=c_+(\kappa,0)\,e^{4(1+i)\kappa^3 t}.$$
The time evolution of the norming constants $c_-(\kappa,t),$ $d_+(\kappa,t),$
and $d_-(\kappa,t)$ appearing in the analogs of (3.4) and (3.6) can be obtained
in a similar way. With the help of (4.3) we obtain (4.7),
and hence the dependency constants $\gamma(\kappa,t)$
and $\eta(\kappa,t)$ appearing in the analogs of (3.3) and (3.5),
respectively, evolve according to (4.8). \qed

\vskip 10 pt
\noindent {\bf 5. A COMPARISON OF
REFERENCES [8] AND [10]}
\vskip 3 pt

As seen from (1.1), if $f(k,x)$ is a solution to (1.1), so are
$f(-k,x),$ $f(ik,x),$ and $f(-ik,x).$ Thus, a solution to (1.1)
known in the region $\Omega$ in the
complex $k$-plane can be extended to the three regions
obtained by rotating $\Omega$ by $\pi/2,$ $\pi,$ and
$3\pi/2,$ respectively, around the
origin of the complex $k$-plane. Moreover,
since the potentials $u$ and $v$ are real valued,
$f(k^*,x)^*$ is also a solution and hence
a solution known in a region in the complex $k$-plane can be
extended to the symmetric region
with respect to the real axis. Thus, solutions
known in $\Omega$ can be extended to the entire
complex $k$-plane.

In [10], some four solutions $\Psi_j(k,x)$ to (1.1) for $j=1,2,3,4$ are presented
on the whole complex $k$-plane
with spatial asymptotics
$$\Psi_j(k,x)=\cases e^{k_jx}[1+o(1)],\qquad x\to+\infty,\\
\noalign{\medskip}
a_j(k)\,e^{k_jx}[1+o(1)],\qquad x\to-\infty,\endcases$$
where the $a_j(k)$ are certain coefficients and the $k_j$ are
as in (2.3).
Comparing the asymptotics as $x\to\pm\infty,$ we see that those four solutions are
related to the Jost and exponential solutions appearing in (2.1) and (2.2) as follows:
$$\Psi_1(k,x)=\displaystyle\frac{\phi_-(k,x)}{A(k)},\qquad
\Psi_2(k,x)=\psi_+(k,x),$$
$$\Psi_3(k,x)=\phi_+(k,x),\qquad
\Psi_4(k,x)=T(k)\psi_-(k,x),$$
where $A$ and $T$ are the coefficients appearing in some of (2.4)-(2.6).
Then, as $k$ moves to the boundary of $\arg k\in(0,\pi/4)$ from the interior,
we see that the reflection coefficients $r_0(k),$ $r_1(k),$ and $r_2(k)$ defined
in [10] are related  as follows to the coefficients used in [8] and in our paper:
$$r_0(k)=R_-(k),\qquad r_1(k)=C_+(k)^*,\qquad r_2(k)=\displaystyle\frac{B_-(k)^*}{A(k)^*}.
$$
Moreover, the quantities $a_j(k)$ appearing in [10] are related to
the quantities used in [8] and in our paper as
$$a_1(k)=\displaystyle\frac{1}{A(k)},
\qquad a_2(k)=\displaystyle\frac{1}{T(k)},
\qquad a_3(k)=A(k),\quad
\qquad a_4(k)=T(k).$$
Thus, the inverse problem has been analyzed in [10] in the special case
$R_\pm(k)=C_\pm(k)=B_\pm(k)=0.$
In that case, $A(k)$ and $T(k)$ simply become
rational functions of $k$ with asymptotics $1+O(1/k)$ as $k\to\infty$
and with appropriate jump conditions on the rays $\arg k=(l-1)\pi/4$ for
$l=1,2,\dots,8.$ One can then formulate the inverse problem on the
entire complex $k$-plane as a Riemann-Hilbert problem and solve it explicitly.

\vskip 10 pt
\noindent {\bf 6. EXAMPLES}
\vskip 3 pt

In this section we present some explicit examples of solutions
and relevant quantities associated with (1.1).
Such examples should help to understand better the scattering
and bound-state data for (1.1). It is already known [5,8,10] that
if $f(k,x)$ is a solution
to the Schr\"odinger equation
$$-f''(k,x)+q(x)\,f(k,x)=k^2\,f(k,x),$$
then $f(k,x)$ is also a solution to
(1.1) when
$$u(x)=-2\,q(x),\qquad v(x)=q(x)^2-q''(x),$$
because in that case we have
$$D^4+DuD+v=(-D^2+q)^2.\tag 6.1$$
Note that (6.1) holds
in our first and fourth examples below, but it does
not hold for our second and third examples.

\noindent {\bf Example 6.1}
Let $u(x)=v(x)=0$ for $x<0$ and
$$u(x)=\displaystyle\frac{4e^{x}}{(1+e^{x})^2},
\qquad v(x)=\displaystyle\frac{2e^{x}(1-e^{x})^2}{(1+e^{x})^4},
\qquad x>0.$$
By using the continuity of $\psi_+,$ $\psi_+',$ $\psi_+''$ at $x=0$ and
the jump condition
$$\psi_+'''(k,0^+)-\psi_+'''(k,0^-)=-u(0^+)\,\psi_+'(k,0),$$
we can determine
all the quantities relevant to (1.1).
In terms of
$$f(k,x):=e^{ikx}\left[1-\displaystyle\frac{2i}{(2k+i)(1+e^{x})}\right],$$
we have the Jost and exponential solutions
$$\psi_+(k,x)=\cases
f(k,x)+C_+(k)\,f(ik,x),\qquad x\ge 0,\\
\noalign{\medskip}
\displaystyle\frac{1}{T(k)}e^{ikx}+\displaystyle\frac{R_+(k)}{T(k)}e^{-ikx}+c_1(k)e^{kx},\qquad x\le 0,
\endcases$$
$$\phi_+(k,x)=\cases
f(ik,x),\qquad x\ge 0,\\
\noalign{\medskip}
A(k)\,e^{-kx}+B_+(k)\,e^{ikx}+c_2(k)e^{kx}+c_3(k)\,e^{-ikx},\qquad x\le 0,
\endcases$$
where
$$A(k)=\displaystyle\frac{(8k^2+1)(2k-1)}{16k^3},\qquad
B_+(k)=\displaystyle\frac{(4-4i)k^2+i}{16k^3(2k+1)},\qquad
c_1(k)=\displaystyle\frac{(2i-2)k+(1+i)}{2k(8k^2+1)},$$
$$
c_2(k)=\displaystyle\frac{-2k+1}{16k^3},\qquad
c_3(k)=\displaystyle\frac{(4+4i)k^2-i}{16k^3(2k+1)},\qquad
R_+(k)=\displaystyle\frac{16k^4+(2+4i)k^2-1}
{(8k^2-i)(16k^4+2ik^2-1)},$$
$$C_+(k)=\displaystyle\frac{(4+4i)k^2+i}{(8k^2+1)(2k+i)(2k-1)},\qquad
T(k)=\displaystyle\frac{2k(2k-1)(2k+1)(2k+i)(8k^2+1)}{(8k^2-i)(16k^4+2ik^2-1)}.$$
The remaining coefficients $R_-,$ $B_-,$ and $C_-$ can easily be evaluated by
using (2.7) and (2.8).
In this example, there is exactly one simple bound state at
$k=\kappa$ with $\kappa:=(1+i)/4,$ where $T(k)$ has a simple pole.
A corresponding bound-state eigenfunction
is a constant multiple of $\psi_+(\kappa,x),$ and it can be chosen as
$$\varphi(\kappa,x)=\cases\displaystyle\frac{e^{-x/4}}{1+e^x}\left\{
\cos(x/4)+\sin(x/4)+e^x\left[ \cos(x/4)-3\sin(x/4)\right]\right\}
,\qquad x\ge 0,\\
\noalign{\medskip}
e^{x/4}\left[ \cos(x/4)-3\sin(x/4)\right],\qquad x\le 0.
\endcases$$
Even though
$A$ and $T$ each have a zero at $k=1/2$ on the ray $\arg k=0,$
their ratio $A/T$
is nonzero at $k=1/2,$ which corresponds to a spectral
singularity and not to a bound state.

\noindent {\bf Example 6.2} Let $u(x)=0$ and $v(x)=-\epsilon \delta(x),$ with
$\delta(x)$ denoting the Dirac delta distribution and $\epsilon$ being a
real, nonzero parameter.
We want $\psi,$ $\psi',$ $\psi''$ to be continuous at $x=0,$ and
$\psi'''(k,0^-)=-\epsilon+\psi'''(k,0^+).$ We find
$$\psi_+(k,x)=\cases
e^{ikx}-\displaystyle\frac{\epsilon}{4k^3+\epsilon}e^{-kx},\qquad x\ge 0,\\
\displaystyle\frac{4k^3+(1-i)\epsilon}{4k^3+\epsilon}e^{ikx}
+\displaystyle\frac{i\epsilon}{4k^3+\epsilon}e^{-ikx}
-\displaystyle\frac{\epsilon}{4k^3+\epsilon}e^{kx},
\qquad x\le 0,\endcases$$
$$\phi_+(k,x)=\cases e^{-kx},\qquad x\ge 0,\\
-\displaystyle\frac{i\epsilon}{4k^3}e^{ikx}
+\displaystyle\frac{i\epsilon}{4k^3}e^{-ikx}
-\displaystyle\frac{\epsilon}{4k^3}e^{kx}+
\displaystyle\frac{4k^3}{4k^3+\epsilon}e^{-kx}
\qquad x\le 0.\endcases$$
The coefficients related to the corresponding scattering problem
are given by
$$\displaystyle\frac{1}{T(k)}=\displaystyle\frac{4k^3+(1-i)\epsilon}{4k^3+\epsilon},\qquad
R_\pm(k)=\displaystyle\frac{i\epsilon}{4k^3+(1-i)\epsilon},$$
$$A(k)=\displaystyle\frac{4k^3+\epsilon}{4k^3},
\qquad B_\pm(k)=-\displaystyle\frac{i\epsilon}{4k^3},
\qquad C_\pm(k)=-\displaystyle\frac{\epsilon}{4k^3+\epsilon}.$$
Note that $A(k)/T(k)$ vanishes on the rays $\arg k=0$ and $\arg k=\pi/4$
only when
$$4k^3+(1-i)\epsilon=0.$$
Since we assume $\epsilon\ne 0,$ we find that there are no such $k$ values
if $\epsilon<0,$ and there exists exactly one $k$ value lying on the ray
$\arg k=\pi/4$ when $\epsilon>0.$ Denoting that $k$-value by $\kappa,$ we
obtain a bound state of multiplicity one at $k=\kappa,$ where
$$\kappa:=\displaystyle\frac{1+i}{2}\root 3\of{\epsilon}.$$
Thus, a bound-state eigenfunction is obtained as
$$\psi_+(\kappa,x)=\cases e^{i\kappa x}+ie^{-\kappa x},
\qquad x\ge 0,\\
\noalign{\medskip}
e^{-i\kappa x}+ie^{\kappa x},
\qquad x\le 0.\endcases$$
Since $\int_{-\infty}^\infty dx\,|\psi_+(\kappa,x)|^2=4/\root 3\of{\epsilon},$
a normalized bound-state eigenfunction is given by
$$\varphi(\kappa,x)=\cases \root 6\of{\epsilon}e^{-\root 3\of{\epsilon}
\,x/2}\left[\cos(
\root 3\of{\epsilon}\,x/2)+\sin(\root 3\of{\epsilon}\,x/2)\right],
\qquad x\ge 0,\\
\noalign{\medskip}
\root 6\of{\epsilon}e^{\root 3\of{\epsilon}\,x/2}\left[\cos(\root 3\of{\epsilon}\,
x/2)-\sin(\root 3\of{\epsilon}\,x/2)\right],
\qquad x\le 0.\endcases$$

\noindent {\bf Example 6.3} For any positive constant $c,$ consider
$$u(x)=\displaystyle\frac{16c^2[1+\sqrt{2}\cosh(2cx)]}{[\sqrt{2}+\cosh(2cx)]^2},$$
$$v(x)=\displaystyle\frac{4c^4[\sqrt{2}\cosh(6cx)-12\cosh(4cx)-5\sqrt{2}\cosh(2cx)+4]}
{[\sqrt{2}+\cosh(2cx)]^4}.$$
We then obtain
$$A(k)=\displaystyle\frac{[k-(1+i)c][k-(1-i)c]}{[k+(1+i)c][k+(1-i)c]},
\qquad T(k)=\displaystyle\frac{[k+(1+i)c][k-(1-i)c]}{[k-(1+i)c][k+(1-i)c]}
,$$
$$R_\pm(k)=0,\qquad C_\pm(k)=0,\qquad B_\pm(k)=0,$$
$$\psi_+(k,x)=e^{ikx}\left[1+\displaystyle\frac{i\alpha(x)}{k+(1+i)c}
+\displaystyle\frac{i\alpha(x)^*}{k-(1-i)c}\right],\qquad \phi_+(k,x)=\psi_+(ik,x),$$
where we have defined
$$\alpha(x):=\displaystyle\frac{-\sqrt{2}\,c-(1+i)c \,e^{-2cx}}{\sqrt{2}+\cosh(2cx)}.$$
This example was presented in [10] in different terminology.
Note that $A(k)/T(k)$ has a double zero at $k=\kappa$
with $\kappa:=(1+i)c,$ which corresponds to a
bound state of multiplicity two. Two linearly independent eigenfunctions are
given by $\psi_+(\kappa,x)$ and $\phi_+(\kappa,x),$ or they can be chosen as
real valued, e.g. as
$$\varphi_1(\kappa,x)=\displaystyle\frac{(\sqrt{2}e^{-cx}+2 e^{cx})\cos(cx)+\sqrt{2}e^{-cx}\sin(cx)}
{\sqrt{2}+\cosh(2cx)},$$
$$\varphi_2(\kappa,x)=\displaystyle\frac{(\sqrt{2}e^{-cx}+2 e^{cx})\sin(cx)-\sqrt{2}e^{-cx}\cos(cx)}
{\sqrt{2}+\cosh(2cx)}.$$

\noindent {\bf Example 6.4}
Consider the potentials
$$u(x)=\displaystyle\frac{-4}{(|x|+1)^2},\qquad v(x)=\displaystyle\frac{-8}{(|x|+1)^4}.$$
Using (2.4)-(2.6) and the continuity of the solutions to (1.1) and the continuity of their
first, second, and third $x$-derivatives, we get
$$T(k)=\displaystyle\frac{k(k^4+k^3+k^2+2k+2)}{(k+i)[k^4+(1+i)k^3+ik^2+(1-i)k+3]},$$
$$A(k)=\displaystyle\frac{(k+1)(k^4+k^3+k^2+2k+2)}{k^5},$$
$$C_+(k)=\displaystyle\frac{-2(k+i)}{k^4+k^3+k^2+2k+2},\qquad B_+(k)=
\displaystyle\frac{-2i(k+1)(k+i)}{k^5},$$
$$R_+(k)=\displaystyle\frac{i(k^2+k+3)}{(k+i)[k^4+(1+i)k^3+ik^2+(1-i)k+3]},$$
$$\psi_+(k,x)=\cases f(k,x)+C_+(k)\,f(ik,x),\qquad x\ge 0\\
\noalign{\medskip}
\displaystyle\frac{1}{T(k)}\,f(-k,-x)+\displaystyle\frac{R_+(k)}{T(k)}
\,f(k,-x)+C_+(k)\,f(ik,-x),\qquad x\le 0,\\
\endcases$$
and for $x\ge 0$ we have $\phi_+(k,x)=f(ik,x),$ while
for $x\le 0$ we have
$$\phi_+(k,x)=A(k)\,f(-ik,-x)+B_+(k)\,f(-k,-x)+
B_+(k)^*\,f(k,-x)-\displaystyle\frac{k^2-2}{k^5}\,f(ik,-x),$$
where we have defined
$$f(k,x):=e^{ikx}\left[1+\displaystyle\frac{i}{k(x+1)}\right].$$
In this example there exists exactly one bound state at
$k=\kappa$ with $\kappa:=0.77\overline{8}(1+i)$
corresponding to the simple pole of
$T(k)$ on the ray $\arg k=\pi/4.$ Note that
we have used an overline to denote the
round-off on the digit. An eigenfunction for
that simple bound state is a constant multiple of
$\psi_+(\kappa,x).$ At
$k=0.47\overline{6}+1.18\overline{3}i$ in the
interior of the sector $\Omega,$
both $A$ and $T$ have simple zeros
without $A/T$ vanishing there; thus, that $k$-value
does not correspond to a bound state and
it corresponds to a nonspectral singularity.

\vskip 10 pt

\noindent{\bf Acknowledgment}.
The research leading to this article
was supported in part by the National Science Foundation under grant
DMS-0610494 and a National Technical University PEBE grant.
The first author is grateful to the colleagues in the
Department of Mathematics at National Technical University of Athens
for their hospitality during his recent visit.

\vskip 10 pt

\noindent {\bf REFERENCES}

\item{[1]} M. J. Ablowitz and P. A. Clarkson, {\it Solitons, nonlinear
evolution equations and inverse scattering,} Cambridge Univ. Press, Cambridge,
1991.

\item{[2]} M. J. Ablowitz and H. Segur, {\it
Solitons and the inverse scattering
transform,} SIAM, Philadelphia, 1981.

\item{[3]} T Aktosun, {\it Solitons and inverse scattering transform,}
In: D. P. Clemence and G. Tang (eds.), {\it Mathematical studies in nonlinear wave propagation,}
Contemp. Math., Vol. {\bf 379}, Amer. Math. Soc., Providence, RI, 2005, pp. 47--62.

\item{[4]} V. Barcilon,
{\it Inverse problem for the vibrating beam in the free-clamped configuration,}
Phil. Transact. Royal Soc. London A {\bf 304}, 211--251 (1982).

\item{[5]} I. M. Gel'fand and L. A. Diki\u\i,
{\it Fractional powers of operators, and Hamiltonian systems,}
Functional Anal. Appl. {\bf 10}, 259--273 (1977).

\item{[6]} H. P. W. Gottlieb,
{\it Isospectral Euler-Bernoulli beams with continuous density and rigidity functions,}
Prof. Royal Soc. London A {\bf 413}, 235--250 (1987).

\item{[7]} J. Hoppe, A. Laptev,
and J. \"Ostensson,
{\it Solitons and the removal of eigenvalues for fourth-order differential operators,}
Int. Math. Res. Not. {\bf 2006}, Art. ID 85050, 14 pp. (2006).

\item{[8]} K. Iwasaki,
{\it Scattering theory for 4th order differential operators. I,}
Japan. J. Math. (N.S.) {\bf 14}, 1--57 (1988).

 \item{[9]} K. Iwasaki,
{\it Scattering theory for 4th order differential operators. II,}
Japan. J. Math. (N.S.) {\bf 14}, 59--96 (1988).

\item{[10]} A. Laptev, R. Shterenberg, V. Sukhanov, and J. \"Ostensson,
{\it Reflectionless potentials for an ordinary differential operator of order four,}
Inverse Problems {\bf 22}, 135--153 (2006).

\item{[11]} S. Novikov, S. V. Manakov, L. P. Pitaevskii, and V. E. Zakharov,
{\it Theory of solitons,}
Consultants Bureau, New York, 1984.

\item{[12]} H. Y. Wang,
{\it Isospectral hierarchies associated with a fourth order eigenvalue problem,}
Ann. Differential Equations {\bf 3}, 359--364 (1987).

\end